\documentclass[prd,%preprint,%onecolumn, 
twocolumn,
aps,showpacs,nofootinbib,nobibnotes,superscriptaddress,preprintnumbers,eqsecnum]{revtex4}
\usepackage{bm}
\usepackage{color}
\usepackage{dcolumn}   
\usepackage{bm}     
\usepackage{bbm}       
\usepackage{amssymb}  
\usepackage{amsmath}
\usepackage{latexsym}
\usepackage{float}
\usepackage{ifthen}
\usepackage{enumerate}
\usepackage{url}
\usepackage{amsopn}

\usepackage{amsfonts}
\usepackage{multirow}
\usepackage{array}
\usepackage{booktabs}
\usepackage{rotating}

\usepackage{ulem}
\newcommand{\ud}{\mathrm{d}} 

\def\clap#1{\hbox to 0pt{\hss#1\hss}}

\def\({\left(}
\def\){\right)}
\def\[{\left[}
\def\]{\right]}
\def\bea{\begin{eqnarray}}
\def\eea{\end{eqnarray}}
\def\be{\begin{equation}}
\def\ee{\end{equation}}
\def\ba{\begin{eqnarray}}
\def\ea{\end{eqnarray}}
\def\beq{\begin{eqnarray}}
\def\eeq{\end{eqnarray}}

\newcommand{\cs}{c_s}

\def\cs{c_{\rm s}}

\def\be{\begin{equation}}
\def\ee{\end{equation}}
\def\ba{\begin{eqnarray}}
\def\ea{\end{eqnarray}}
\def\beq{\begin{eqnarray}}
\def\eeq{\end{eqnarray}}

\def\L*{{\cal L}_*}
\def\L{\mathcal{L}}
\def\({\left(}
\def\){\right)}

\def\<{\langle}
\def\>{\rangle}

\def\cs2{c_{s}^{2}}

\def\be{\begin{equation}}
\def\ee{\end{equation}}
\def\ba{\begin{eqnarray}}
\def\ea{\end{eqnarray}}
\def\beq{\begin{eqnarray}}
\def\eeq{\end{eqnarray}}

\def\L*{{\cal L}_*}
\def\L{\mathcal{L}}
\def\({\left(}
\def\){\right)}

\def\<{\langle}
\def\>{\rangle}

\allowdisplaybreaks

\begin{document}
%\hspace{5.2in} \mbox{NORDITA-2015-38}\\\vspace{1.53cm} % Preprint number

\title{Dipolar Dark Matter as an Effective Field Theory}

\date{\today}

\author{Luc Blanchet} \email{luc.blanchet@iap.fr}
\affiliation{$\mathcal{G}\mathbb{R}\varepsilon{\mathbb{C}}\mathcal{O}$
  Institut d'Astrophysique de Paris --- UMR 7095 du CNRS,
  \ Universit\'e Pierre \& Marie Curie, 98\textsuperscript{bis}
  boulevard Arago, 75014 Paris, France}

\author{Lavinia Heisenberg} \email{lavinia.heisenberg@eth-its.ethz.ch}
\affiliation{Institute for Theoretical Studies, ETH Zurich, 
\\ Clausiusstrasse 47, 8092 Zurich, Switzerland}

\date{\today}

\begin{abstract}
Dipolar Dark Matter (DDM) is an alternative model motivated by the
challenges faced by the standard cold dark matter model to describe
the right phenomenology at galactic scales. A promising realisation of
DDM was recently proposed in the context of massive bigravity
theory. The model contains dark matter particles, as well as a vector
field coupled to the effective composite metric of bigravity. This
model is completely safe in the gravitational sector thanks to the
underlying properties of massive bigravity. In this work we
investigate the exact decoupling limit of the theory, including the
contribution of the matter sector, and prove that it is free of ghosts
in this limit. We conclude that the theory is acceptable as an
Effective Field Theory below the strong coupling scale.
\end{abstract}

%PACS NEEDED

\maketitle

%%%%%%%%%%%%%%%%%%%%%%%%%%%%%%%%%%%%%%%%%%%%%%%%%%%%%%%%%%%%%%%%%%%%%%%%%
%%%%%%%%%%%%%%%%%%%%%%%%%%%%%%%%%%%%%%%%%%%%%%%%%%%%%%%%%%%%%%%%%%%%%%%%%
\section{Introduction}
\label{sec:intro}

We are witnesses of centenaries. The year 2015 marked the 100th
anniversary of Albert Einstein's elaborate theory of General
Relativity (GR), while 2016 celebrated the centenary of the first
paper on gravitational waves by the announcement of their experimental
detection~\cite{GW150914}. GR meets the requirements of the underlying
physics in a broad range of scales, from black hole to solar system
size. It stood up to intense scrutiny and prevailed against all
alternative competitors. It constitutes the bedrock upon which our
fundamental understanding of gravity relies. However,
%like everything else, it is not quite perfect. 
some important questions remain. 

The lack of renormalizability motivates the modifications of gravity
in the ultraviolet (UV), that incorporate the quantum nature of
gravity. The singularities present in the classical theory could be
regularized by the new physics~\cite{Jimenez:2014fla}. The UV
modifications might also dictate a different scenario for the early
Universe as an alternative to inflation~\cite{Jimenez:2015jqa}. The
inflaton field in the standard picture might be just a reminiscent of
the modification of gravity in the UV.

From a more observational point of view, GR faces additional
challenges on cosmological scales. In order to account for the
observed amount of ingredients of the Universe, it is necessary to
introduce dark matter and dark energy despite of their unclear
origin. Notwithstanding of remarkable efforts, the dark matter has so
far not been directly detected. Concerning the dark energy,
%component of the Universe, 
the standard model in form of a cosmological constant $\Lambda$
accounts for most of the observations even though it faces the
unnaturalness problem~\cite{Weinberg:1988cp}. Combined with the
non-baryonic cold dark matter (CDM) component, the model explains
remarkably well the observed fluctuations of the cosmic microwave
background and the formation of large scale structures.

Albeit the many successes of the $\Lambda$-CDM model at large scales,
it has difficulties to explain the observations of dark matter at
galactic scales. For instance, it is not able to account for the tight
correlations between dark and luminous matter in galaxy
halos~\cite{SandMcG02, FamMcG12}. In this remark, the first
unsatisfactory discrepancy comes from the observed Tully-Fisher
relation between the baryonic mass of spiral galaxies and their
asymptotic rotation velocity. Another discrepancy, perhaps more
fundamental, comes from the correlation between the
presence of dark matter and the acceleration scale~\cite{McG00,
  McG11}. The prevailing view regarding these problems is that they
should be resolved once we understand the baryonic processes that
affect galaxy formation and evolution~\cite{SilkM12}. However, this
explanation is challenged by the fact that galactic data are in
excellent agreement with the MOND (MOdified Newtonian Dynamics)
empirical formula~\cite{Milg1, Milg2, Milg3}. From a phenomenological
point of view, this formula accommodates remarkably well all
observations at galactic scales. Unfortunately, extrapolation of the
MOND formula to the larger scale of galaxy clusters confronts an
incorrect dark matter distribution~\cite{GD92, PSilk05, Clowe06,
  Ang08, Ang09}.

The ideal scenario would be to have a hybrid model in which the
properties of the $\Lambda$-CDM model are naturally incorporated on
large scales, whereas the MOND formula would take place on galactic
scales. There have been many attempts to embed the physics beyond the
MOND formula into an approved relativistic theory, either \textit{via}
invoking new propagating fields without dark matter~\cite{Sand97,
  Bek04, Sand05, ZFS07, Halle08, bimond1, BDgef11}, or by considering
MOND as an emergent phenomenology~\cite{B07mond, BLSF09, BL08, BL09,
  Sand11, BM11, Bernard:2014psa, K15, BK15, V16}.

Here we consider a model of the latter class, called Dipolar Dark
Matter (DDM)~\cite{BL08, BL09, Bernard:2014psa}. The most compelling
version of DDM has been recently developed, based on the formalism of
massive bigravity theory~\cite{Blanchet:2015sra, Blanchet:2015bia}. To
describe the potential interactions between the two metrics of
bigravity the model uses the effective composite metric introduced in
Refs.~\cite{dRHRa, dRHRb, Heisenberg:2014rka}. Two species of dark
matter particles are separately coupled to the two metrics, and an
internal vector field that links the two dark matter species is
coupled to the effective composite metric. The MOND formula is
recovered from a mechanism of gravitational polarization in the non
relativistic approximation. The model has the potential to
reproduce the physics of the $\Lambda$-CDM model at large
cosmological scales.

In the present paper we address the problem of whether there are ghost
instabilities in this model. The model itself~\cite{Blanchet:2015sra,
  Blanchet:2015bia} will be reviewed in Sec.~\ref{sec:DDM}. The model is safe 
 in the gravitational sector because it uses the ghost-free framework of massive bigravity. 
 The interactions of the matter fields with the effective metric reintroduce a ghost
in the matter sector beyond the strong coupling scale, as found in~\cite{dRHRa, dRHRb}. 
In our model, apart from this effective coupling the different
  species of matter fields interact with each other via an internal vector field.
  This additional coupling might spoil the property of ghost freedom within the strong
  coupling scale.
  We therefore investigate, in Sec.~\ref{sec:DL},
the exact decoupling limit (DL) of our model, crucially including the
contributions coming from the matter sector and notably from the
internal vector field. The model dictates what are the relevant
scalings of the matter fields in terms of the Planck mass in the
DL. Using that, we shall prove that the theory is free of ghosts in the
DL and conclude that it is acceptable as an Effective Field Theory
below the strong coupling scale. We end the paper with a few
concluding remarks in Sec.~\ref{sec:concl}.

%%%%%%%%%%%%%%%%%%%%%%%%%%%%%%%%
%%%%%%%%%%%%%%%%%%%%%%%%%%%%%%%%
\section{Dipolar Dark Matter}
\label{sec:DDM}

The model that we would like to study in this work is the dark matter
model proposed in Ref.~\cite{Blanchet:2015sra} where the Dipolar Dark
Matter (DDM) at small galactic scales is connected to bimetric gravity
based on the ghost-free bimetric formulation of massive
gravity~\cite{dRGT10, Hassan:2011zd}. The action of a successful
realisation was investigated in~\cite{Blanchet:2015bia} and we would
like to push forward the analysis performed there. The Lagrangian is
the sum of a gravitational part, based on massive bigravity theory,
plus a matter part: $\mathcal{L} = \mathcal{L}_\text{grav} +
\mathcal{L}_\text{mat}$. The gravitational part reads
\begin{align}
\mathcal{L}_\text{grav} = \frac{M_g^2}{2}\sqrt{-g} \,R_g +
\frac{M_f^2}{2} \sqrt{-f} \,R_f + m^2M_\text{eff}^2
\,\sqrt{-g_\text{eff}}\,,\label{Lgrav}
\end{align}
where $R_g$ and $R_f$ denote the Ricci scalars of the two metrics
$g_{\mu\nu}$ and $f_{\mu\nu}$, with the corresponding Planck scales
$M_g$ and $M_f$ and the interactions carrying another Planck scale
$M_\text{eff}$, together with the graviton's mass $m$. In this
formulation, the ghost-free potential interactions between the two
metrics are defined as the square root of the determinant of the
effective composite metric~\cite{dRHRa, dRHRb, Heisenberg:2014rka}
\begin{equation}\label{effmetric}
g^\text{eff}_{\mu\nu}=\alpha^2 g_{\mu\nu} +2\alpha\beta
\,\mathcal{G}^\text{eff}_{\mu\nu} +\beta^2 f_{\mu\nu}\,,
\end{equation}
with the arbitrary dimensionless parameters $\alpha$ and $\beta$
(typically of the order of one). Here
$\mathcal{G}^\text{eff}_{\mu\nu}$ denotes the effective metric in the
previous DDM model~\cite{Bernard:2014psa}, given by
$\mathcal{G}^\text{eff}_{\mu\nu} = g_{\mu\rho}X^\rho_\nu$ where
$X=\sqrt{g^{-1}f}$, or equivalently $\mathcal{G}^\text{eff}_{\mu\nu} =
f_{\mu\rho}Y^\rho_\nu$ where $Y=\sqrt{f^{-1}g}$. It is trivial to see
that the square root of the determinant of this effective metric
$g^\text{eff}_{\mu\nu}$ corresponds to the allowed ghost-free
potential interactions~\cite{dRHRa}.

The matter part of the model will consist of ordinary baryonic matter
and a dark sector including dark matter particles. The crucial feature
of the model is the presence of a vector field $\mathcal{A}_\mu$ in
the dark sector, that is sourced by the mass currents of dark matter
particles and represents a ``graviphoton''~\cite{SCHERK1979265}. This
vector field stabilizes the DDM medium and ensures a mechanism of
``gravitational polarisation''. The matter action reads
\begin{align}\label{Lmat}
\mathcal{L}_\text{mat} =& -
\sqrt{-g}\bigl(\rho_\text{bar}+\rho_g\bigr) - \sqrt{-f} \,\rho_f
\nonumber\\ & + \sqrt{-g_\text{eff}} \biggl[
  \mathcal{A}_\mu\bigl(j_g^\mu-j_f^\mu\bigr) + \lambda
  M_\text{eff}^2\,\mathcal{W}\bigl(\mathcal{X}\bigr) \biggr]\,.
\end{align}
Note the presence of a non-canonical kinetic term for the vector field
in form of a function $\mathcal{W}(\mathcal{X})$ of
\begin{equation}\label{X}
\mathcal{X} = -
\frac{\mathcal{F}^{\mu\nu}\mathcal{F}_{\mu\nu}}{4\lambda}\,,
\end{equation}
with the field strength defined by $\mathcal{F}^{\mu\nu} \equiv
g_\text{eff}^{\mu\rho}
g_\text{eff}^{\nu\sigma}\mathcal{F}_{\rho\sigma}$ where
$\mathcal{F}_{\mu\nu} = \partial_\mu\mathcal{A}_\nu -
\partial_\nu\mathcal{A}_\mu$. The form of the function
$\mathcal{W}(\mathcal{X})$ has been determined by demanding that the
model reproduces the MOND phenomenology at galactic
scales~\cite{Bernard:2014psa, Blanchet:2015bia, Bernard:2015gwa}. This
corresponds to the limit $\mathcal{X}\to 0$ and we have
\begin{equation}\label{W(X)}
\mathcal{W}(\mathcal{X})=\mathcal{X}-\frac23 (\alpha+\beta)^2
\mathcal{X}^{3/2}+\mathcal{O}\left(\mathcal{X}^2\right) \,,
\end{equation}
so that the leading term in the action~\eqref{Lmat} is
\begin{equation}\label{W(X)lead}
\lambda M_\text{eff}^2\,\mathcal{W}\bigl(\mathcal{X}\bigr) = -
\frac{M_\text{eff}^2}{4} \mathcal{F}^{\mu\nu}\mathcal{F}_{\mu\nu} +
\mathcal{O}\left(\mathcal{F}^{3}\right) \,.
\end{equation}
Hence, we observe that the coupling scale of the vector field is
dictated by $M_\text{eff}$, while the parameter $\lambda$ enters into
higher-order corrections. In order to recover the correct MOND regime
for very weak accelerations of baryons in the ordinary $g$ sector,
\textit{i.e.} below the MOND acceleration scale $a_0$, these constants
have been determined as~\cite{Blanchet:2015bia}\footnote{Recall also
  that the MOND acceleration $a_0$ is of the order of the cosmological
  parameters, and thus is extremely small in Planck units, $a_0\sim
  10^{-63}\,M_\text{Pl}$.}
\begin{equation}\label{Mefflambda}
M_\text{eff}=\sqrt{2} \,r_g
M_\text{Pl}\quad\text{and}\quad\lambda=\frac{a_0^2}{2}\,.
\end{equation}
Here $M_\text{Pl}$ represents the standard Planck constant of GR and
the constant $r_g$ is defined below. It is worth mentioning that the
standard Newtonian limit in the ordinary $g$ sector is obtained by
imposing the relation
\begin{equation}
M_g^2+\frac{\alpha^2}{\beta^2}M_f^2= M_\text{Pl}^2\,.
\end{equation}
Thus, in this model the three mass scales $M_g$, $M_f$ and
$M_\text{eff}$ are of the order of the Planck mass.

We represent the scalar energy densities of the ordinary pressureless
baryons, and the two species of pressureless dark matter particles by
$\rho_\text{bar}$, $\rho_g$ and $\rho_f$ respectively. Such densities
are conserved in the usual way with respect to their respective
metrics, hence $\nabla^g_\mu(\rho_\text{bar} u_\text{bar}^\mu)=0$,
$\nabla^g_\mu(\rho_g u_g^\mu)=0$ and $\nabla^f_\mu(\rho_f u_f^\mu)=0$,
with the four velocities being normalized as
$g_{\mu\nu}u_\text{bar}^\mu u_\text{bar}^\nu=-1$, $g_{\mu\nu}u_g^\mu
u_g^\nu=-1$ and $f_{\mu\nu}u_f^\mu u_f^\nu=-1$. The respective
stress-energy tensors are defined as
$T_\text{bar}^{\mu\nu}=\rho_\text{bar} u_\text{bar}^\mu
u_\text{bar}^\nu$, $T_g^{\mu\nu}=\rho_g u_g^\mu u_g^\nu$ and
$T_f^{\mu\nu}=\rho_f u_f^\mu u_f^\nu$. The pressureless baryonic fluid
obeys the geodesic law of motion $a^\text{bar}_\mu \equiv
u_\text{bar}^\nu \nabla^{g}_\nu u^\text{bar}_\mu=0$, hence
$\nabla_g^\nu T^\text{bar}_{\mu\nu}=0$. On the other hand, because of
their coupling to the vector field, the dark matter fluids pursue a
non-geodesic motion:
\begin{subequations}\label{EOMdm}
\begin{align}
\nabla_g^\nu T^g_{\mu\nu} &= J_g^\nu
\mathcal{F}_{\mu\nu}\,,\\ \nabla_f^\nu T^f_{\mu\nu} &= - J_f^\nu
\mathcal{F}_{\mu\nu}\,,\label{EOMdmf}
\end{align}
\end{subequations}
where the dark matter currents $J_g^\mu$ and $J_f^\mu$ are related to
those appearing in Eq.~\eqref{Lmat} by
\begin{equation}\label{Jj}
J_g^\mu =
\frac{\sqrt{-g_\text{eff}}}{\sqrt{-g}}\,j_g^\mu\quad\text{and}\quad
J_f^\mu = \frac{\sqrt{-g_\text{eff}}}{\sqrt{-f}}\,j_f^\mu\,.
\end{equation}
It remains to specify the link between these currents and the scalar
densities $\rho_g$ and $\rho_f$ of the particles. This is provided by
$J_g^\mu=r_g \rho_g u_g^\mu$ and $J_f^\mu=r_f \rho_f u_f^\mu$, where
$r_g$ and $r_f$ are two constants of the order of one, which can be
interpreted as the ratios between the ``charge'' of the particles
(with respect to the vector interaction) and their inertial mass. For
correctly recovering MOND we must have $\alpha r_g=\beta
r_f$~\cite{Blanchet:2015bia}.

Whereas, the stress-energy tensor of the vector field
$\mathcal{A}_\mu$ is obtained by varying~\eqref{Lmat} with respect to
$g_{\mu\nu}^\text{eff}$ (holding the $g$ and $f$ metrics fixed) and
corresponds to
\begin{equation}\label{Tgeff}
T_{g_{\text{eff}}}^{\mu\nu} = M_\text{eff}^2\Bigl[\mathcal{W}_{\mathcal{X}}
  \,\mathcal{F}^{\mu\rho}\mathcal{F}^\nu_{\phantom{\nu}\rho} +
  \lambda\mathcal{W}\,g_\text{eff}^{\mu\nu}\Bigr]\,,
\end{equation}
where $\mathcal{W}_{\mathcal{X}}\equiv\ud\mathcal{W}/\ud\mathcal{X}$.
The evolution of the vector field is dictated by the Maxwell law
\begin{equation}\label{eqAmu}
\nabla^{g_\text{eff}}_\nu\Bigl[
  \mathcal{W}_{\mathcal{X}}\mathcal{F}^{\mu\nu}\Bigr] =
\frac{1}{M_\text{eff}^2}\bigl(j^\mu_g-j^\mu_f\bigr)\,,
\end{equation}
where the covariant derivative associated with $g_\text{eff}$ is
denoted by $\nabla^{g_\text{eff}}_\nu$. Together with the conservation
of the currents, $\nabla^{g_\text{eff}}_\mu j_g^\mu=0$ and
$\nabla^{g_\text{eff}}_\mu j_f^\mu=0$, the equations of motion for the
vector field can also be expressed as
\begin{equation}\label{divtaueff}
\nabla_{g_\text{eff}}^\nu T^{g_\text{eff}}_{\mu\nu} = -
\bigl(j^\nu_g-j^\nu_f \bigr)\mathcal{F}_{\mu\nu}\,,
\end{equation}
and we can combine these equations of motion all together into a
``global'' conservation law
\begin{equation}\label{conslaw}
\sqrt{-g_\text{eff}}\,\nabla_{g_\text{eff}}^\nu
T^{g_\text{eff}}_{\mu\nu} + \sqrt{-g}\,\nabla_g^\nu T^g_{\mu\nu} +
\sqrt{-f}\,\nabla_f^\nu T^f_{\mu\nu} = 0\,.
\end{equation}
%
%Finally it remains to specify the link between the dark matter
%currents $J_g^\mu$ and $J_f^\mu$ (or equivalently $j_g^\mu$ and
%$j_f^\mu$) and their mass density. This is provided by $J_g^\mu=r_g
%\rho_g u_g^\mu$ and $J_f^\mu=r_f \rho_f u_f^\mu$, where the two
%constants $r_g$ and $r_f$ specify the ratios between the charge (with
%respect to the vector interaction $\mathcal{A}_\mu$) and the inertial
%mass of the dark matter particles.

%In Ref.~\cite{Blanchet:2015bia} all the parameters of the model
%(notably the masses $M_g$, $M_f$, $M_\text{eff}$, the parameter
%$\lambda$, the constants $r_g$ and $r_f$ and the constants $\alpha$
%and $\beta$ entering the composite metric~\eqref{effmetric}), as well
%as the precise form of the function $\mathcal{W}(\mathcal{X})$, have
%been determined so that the model in order reproduces the MOND
%phenomenology at galactic scales~\cite{Blanchet:2015bia,
%  Bernard:2015gwa}. Notably, the parameter $\lambda$ was then related
%to the MOND acceleration scale $a_0$. In the present paper we shall
%rather work with the general model~\eqref{Lgrav}--\eqref{Lmat}.

%%%%%%%%%%%%%%%%%%%%%%%%%%%%%%%%
%%%%%%%%%%%%%%%%%%%%%%%%%%%%%%%%
\section{Decoupling Limit}
\label{sec:DL}

Being based on massive bigravity theory, the gravitational sector of
the model, Eq.~\eqref{Lgrav}, is ghost-free up to any order in
perturbation theory~\cite{dRGT10, Hassan:2011zd}. In addition, the
baryonic and dark matter particles can be coupled separately to either
the $g$ metric or $f$ metric without changing this
property~\cite{dRHRa}. The case of the pure matter coupling between
the vector field $\mathcal{A}_\mu$ and the effective composite metric
$g_\text{eff}$ in Eqs.~\eqref{Lmat}--\eqref{X}, is not trivial. In
that case, it was shown in Ref.~\cite{dRHRa} that the coupling is
ghost-free in the mini-superspace and in the decoupling
limit. Furthermore it is known that such coupling to the composite
metric is unique in the sense that it is the only non-minimal matter
coupling that maintains ghost-freedom in the decoupling
limit~\cite{deRham:2015cha, Huang:2015yga, Heisenberg:2015iqa}.

\begin{figure}[t]
\begin{center}
\includegraphics[width=9cm]{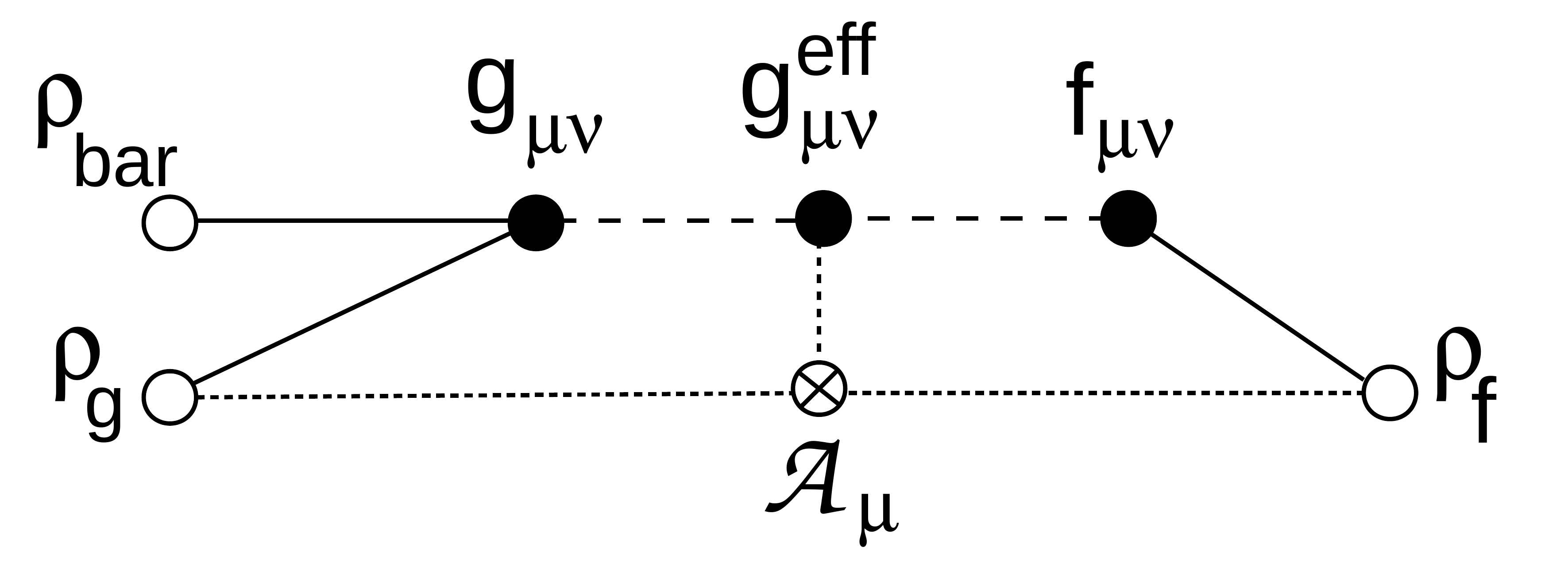}
\end{center}
\caption{Schematic structure of the model. The two metrics of
  bigravity $g_{\mu\nu}$ and $f_{\mu\nu}$ interact through the
  effective composite metric $g^\text{eff}_{\mu\nu}$, but also
  indirectly, \textit{via} the particles $\rho_g$ and $\rho_f$ and the
  vector field $\mathcal{A}_\mu$.}\label{fig}
\end{figure}
However, in our model the vector field is also coupled to the $g$ and
$f$ particles, through the standard interaction term
$\propto\mathcal{A}_\mu(j_g^\mu-j_f^\mu)$. This term plays a crucial
role for the dark matter model to work. This coupling introduces a
suplementary, indirect interaction between the two metrics of
bigravity, \textit{via} the $g$ and $f$ particles coupled together by
the term $\mathcal{A}_\mu(j_g^\mu-j_f^\mu)$. See Fig.~\ref{fig} for a
schematic illustration of the interactions in the model. As a result
it was found in Ref.~\cite{Blanchet:2015bia} that a ghost is
reintroduced in the dark matter sector in the full theory. The aim of
this paper is to investigate the occurence and mass of this ghost, and
whether or not the decoupling limit (DL) is maintained ghost-free. If
the latter is true, then the model can be used in a consistent way as
an Effective Field Theory valid below the strong coupling scale.

We now detail the analysis of the DL interactions in the graviton and
matter sectors. We follow the preliminary work~\cite{Blanchet:2015bia}
and investigate the scale of the reintroduced Boulware-Deser (BD)
ghost~\cite{Boulware:1973my}. We first decouple the interactions below
the strong coupling scale from those entering above it, and
concentrate on the pure interactions of the helicity-0 mode of the
massive graviton. Using the St\"uckelberg trick, we restore the broken
gauge invariance in the $f$ metric by replacing it by
\begin{equation}\label{stuck1}
\tilde{f}_{\mu\nu} = f_{ab}\partial_\mu \phi^a \partial_\nu \phi^b\,,
\end{equation}
where $a, b=0,1,2,3$ and the four St\"uckelberg fields $\phi^a$ are
decomposed into the helicity-0 mode $\pi$ and the helicity-1 mode
$A^a$,
\begin{equation}\label{stuck2}
\phi^a=x^a-\frac{m A^a}{\Lambda^3_3} -\frac{f^{ab}\partial_b
  \pi}{\Lambda^3_3}\,.
\end{equation}
Here $\Lambda_3 \equiv (m^2 M_\text{eff})^{1/3}$ denotes the strong
coupling scale. Note, that we define it with respect to $M_\text{eff}$
here, since the potential interactions scale with $m^2M_\text{eff}^2$
in our case.  

It is well known that the would-be BD ghost in the DL, would come in
the form of higher derivative interactions of the helicity-0 mode at
the level of the equations of motion. Therefore we shall only follow
the contributions of the helicity-0 mode $\pi$ and neglect the
interactions of the helicity-1 mode $A^a$. For simplicity we do not
write the tilde symbol over the St\"ukelbergized version of the
metric~\eqref{stuck1}. Thus, considering also the helicity-2 mode in
the $g$ metric, we have\footnote{If there is a BD ghost in the DL,
  then it will manifest itself in the higher-order equations of motion
  of the helicity-0 mode. For this purpose, it will be enough to
  follow closely the contributions of the matter couplings to the
  helicity-0 mode equations of motion and decouple the dynamics of the
  secondary helicity-2 mode in $f$. The contributions of the latter,
  as derived in~\cite{Fasiello:2013woa}, will not play any role in our
  analysis and will not change the final conclusions. The same is true
  for the contributions of the helicity-1 mode.}
\begin{subequations}\label{eq:fDL}
\begin{align}
g_{\mu\nu} &= \biggl(\eta_{\mu\nu}+\frac{h_{\mu\nu}}{M_g}\biggr)^2\,,
\\ f_{\mu\nu} &=
\biggl(\eta_{\mu\nu}-\frac{\Pi_{\mu\nu}}{\Lambda^3_3}\biggr)^2\,,
\end{align}
\end{subequations}
where we introduced the notation $\Pi_{\mu\nu} \equiv \partial_\mu
\partial_\nu \pi$ for convenience, and raised and lowered indices with
the Minkowski metric $\eta_{\mu\nu}$. The effective metric reads then
\begin{equation}\label{effective}
g_{\mu\nu}^{\text{eff}} = \Bigl((\alpha+\beta)\eta_{\mu\nu} +
K_{\mu\nu} \Bigr)^2\,,
\end{equation}
in which we have introduced as a short-cut notation the linear combination
\begin{equation}\label{kmatrix}
 K_{\mu\nu} = \frac{\alpha}{M_g}h_{\mu\nu} -
 \frac{\beta}{\Lambda^3_3}\Pi_{\mu\nu} \,.
 \end{equation}
We will as next investigate the different contributions in the
gravitational and matter sectors.

\subsection{Gravitational sector}

There is no contribution of the Einstein-Hilbert term to the
helicity-0 mode, since this is invariant under diffeomorphisms. On the
other hand, there will be different contributions coming from the
ghost-free potential interactions. The allowed potential interactions
between the metrics $g$ and $f$ have been chosen in our model to be
given by the square root of the determinant of the composite
metric~\eqref{effective}, which becomes in this case
\begin{equation}\label{detgeff}
\sqrt{-g_\text{eff}} = \sum_{n=0}^4 (\alpha+\beta)^{4-n} e_{(n)}(K)\,,
\end{equation}
where $e_{(n)}(K)$ denote the usual symmetric polynomials associated
with the matrix $K_\mu^\rho\equiv\eta^{\rho\nu}K_{\mu\nu}$, and given
by products of antisymmetric Levi-Cevita tensors,
\begin{subequations}\label{polynomialseps}
\begin{align}
e_{(0)}(K) &= - \frac{1}{24}
\varepsilon^{\mu\nu\rho\sigma}\varepsilon_{\mu\nu\rho\sigma}\,,
\\ e_{(1)}(K) &= - \frac{1}{6}
\varepsilon^{\mu\nu\rho\sigma}\varepsilon_{\mu\nu\rho\lambda}
K^\lambda_\sigma\,, \\ e_{(2)}(K) &= - \frac{1}{4}
\varepsilon^{\mu\nu\rho\sigma}\varepsilon_{\mu\nu\tau\lambda}
K^\tau_\rho K^\lambda_\sigma\,, \\ e_{(3)}(K) &= - \frac{1}{6}
\varepsilon^{\mu\nu\rho\sigma}\varepsilon_{\mu\pi\tau\lambda}
K^\pi_\nu K^\tau_\rho K^\lambda_\sigma\,, \\ e_{(4)}(K) &= - \frac{1}{24}
\varepsilon^{\mu\nu\rho\sigma}\varepsilon_{\epsilon\pi\tau\lambda}
K^\epsilon_\mu K^\pi_\nu K^\tau_\rho K^\lambda_\sigma\,.
\end{align}
\end{subequations}
In particular, we see that $e_{(4)}(K)=\text{det}(K)$.

First of all, the pure helicity-0 mode in the ghost-free potential
interactions~\eqref{detgeff} will come in the form of total
derivatives~\cite{deRham10, dRGT10}. Indeed, as is clear from their
definitions~\eqref{polynomialseps} in terms of antisymmetric
Levi-Cevita tensors, the symmetric polynomials
$e_{(n)}(\Pi)\equiv\mathcal{L}^{\text{der}}_{(n)}(\Pi)$ fully encode
the total derivatives at that order, and thus will not contribute to
the equation of motion of the helicity-0 mode. In fact, in
Ref.~\cite{deRham10}, this very same property of total derivatives of
the leading contributions at each order was used to build the
ghost-free interactions away from $h=0$. Secondly, there will be the
pure interactions of the helicity-2 mode, obtained by setting $\Pi=0$,
and these will come with the corresponding inverse powers of
$M_g$. Finally, there will be the mixed interactions between the
helicity-2 and helicity-0 modes.

We are after the leading interactions in the DL, which correspond to
sending all the Planck scales to infinity,
\begin{equation}\label{scalingDL1}
M_\text{Pl} \to \infty\,, ~~M_{g} \to \infty\,, ~~M_\text{eff} \to
\infty\,, ~~M_f \to \infty \,,
\end{equation}
together with the graviton's mass $m \to0$, while keeping
\begin{equation}\label{scalingDL2}
\biggl\{\Lambda^3_3 = m^2 M_\text{eff}\,, ~~\frac{M_g}{M_\text{Pl}}\,,
~~\frac{M_\text{eff}}{M_\text{Pl}}\,, ~~\frac{M_f}{M_\text{Pl}}
\biggr\} = \text{const}\,.
\end{equation}
Taking into account the factor $m^2 M_\text{eff}^2$ in front of the
potential interactions, one immediately observes that the pure
non-linear interactions of the helicity-2 modes do not contribute to
the DL. As we already mentioned, the pure helicity-0 mode interactions
do not contribute either. So it remains the mixed terms, for which the
only surviving terms will be linear in the helicity-2 mode, and we
finally obtain
\begin{equation}\label{potintDL}
m^2 M_\text{eff}^2 \sqrt{-g_\text{eff}} = \sum_{n=1}^3
\frac{a_n}{\Lambda_3^{3(n-1)}} h^{\mu\nu} P_{\mu\nu}^{(n)}(\Pi) +
\mathcal{O}\left(\frac{1}{M_g}\right)\,,
\end{equation}
where $a_n\equiv
(\frac{M_\text{eff}}{M_g})^{n+1}\alpha(-\beta)^n(\alpha+\beta)^{3-n}$
and we posed
\begin{equation}
P_{\mu\nu}^{(n-1)}(\Pi) \equiv \frac{\partial e_{(n)}(\Pi)}{\partial
  \Pi^{\mu\nu}}\,.
\end{equation} 
In arriving at Eq.~\eqref{potintDL} we have removed the trivial
constant term in~\eqref{detgeff}, and ignored the ``tadpole'' which is
simply proportional to the trace $[h]=h^{\mu}_{\mu}$ and can be
eliminated by choosing an appropriate de Sitter background (see,
\textit{e.g.}, a discussion in~\cite{Blanchet:2015bia}).

We can then write the total contribution of the gravitational sector
in the DL, including that coming from the Einstein-Hilbert term of the
$g$ metric, which enters only at the leading quadratic order in
$h_{\mu\nu}$,
\begin{equation}\label{LgravDL}
\mathcal{L}^\text{DL}_\text{grav} = -
h^{\mu\nu}\mathcal{E}^{\rho\sigma}_{\mu\nu}h_{\rho\sigma} +
\sum_{n=1}^3 \frac{a_n}{\Lambda_3^{3(n-1)}} h^{\mu\nu}
P_{\mu\nu}^{(n)}(\Pi)\,,
\end{equation}
where $\mathcal{E}^{\rho\sigma}_{\mu\nu}$ is the usual Lichnerowicz
operator on a flat background as defined by
\begin{align}\label{lichne}
-2\mathcal{E}_{\mu\nu}^{\rho\sigma}h_{\rho\sigma} &=
\Box\bigl(h_{\mu\nu}-\eta_{\mu\nu}h\bigr) + \partial_\mu\partial_\nu h
\nonumber\\ &- 2 \partial_{(\mu} H_{\nu)} +
\eta_{\mu\nu}\partial_{\rho} H^{\rho}\,,
\end{align}
with $h=[h]=h^\mu_\mu$ and $H_\mu=\partial_\nu
h^\nu_\mu$. \vspace{-0.1cm} The symmetric tensors $P^{(n)}_{\mu\nu}$
are conserved, \textit{i.e.}  $\partial_\nu P_{(n)}^{\mu\nu}=0$. For
an easier comparison with the literature we give them as the product
of two Levi-Cevita tensors appropriately contracted with the second
derivative of the helicity-0 field,
\begin{subequations}\label{Pn}\begin{align}
P^{(1)}_{\mu\nu}(\Pi) &= -\frac{1}{2}
\varepsilon_{\mu}^{\phantom{\mu}\lambda\rho\sigma}\varepsilon_{\nu\lambda\rho\tau}
\Pi^{\tau}_{\sigma}\,,\\ P^{(2)}_{\mu\nu}(\Pi) &= -\frac{1}{2}
\varepsilon_{\mu}^{\phantom{\mu}\lambda\rho\sigma}\varepsilon_{\nu\lambda\pi\tau}
\Pi^{\pi}_{\rho}\Pi^{\tau}_{\sigma}\,,\\ P^{(3)}_{\mu\nu}(\Pi) &=
-\frac{1}{6}
\varepsilon_{\mu}^{\phantom{\mu}\lambda\rho\sigma}\varepsilon_{\nu\epsilon\pi\tau}
\Pi^{\epsilon}_{\lambda}\Pi^{\pi}_{\rho}\Pi^{\tau}_{\sigma}\,.
\end{align}
\end{subequations}

The first two interactions between the helicity-0 and helicity-2
fields in the Lagrangian~\eqref{LgravDL} can be removed by the change
of variable, defining
\begin{equation}\label{fieldredef}
\hat{h}_{\mu \nu} \equiv h_{\mu\nu} - \frac{a_1}{2} \pi \,\eta_{\mu
  \nu} + \frac{a_2}{2\Lambda_3^3} \partial_\mu\pi \partial_\nu \pi\,.
\end{equation}
In this way the Lagrangian of the gravitational sector in the
decoupling limit becomes~\cite{deRham10}
\begin{align}\label{LgravDLhat}
\mathcal{L}^\text{DL}_\text{grav} =& -
\hat{h}^{\mu\nu}\mathcal{E}^{\rho\sigma}_{\mu\nu}\hat{h}_{\rho\sigma}
+ \sum_{n=0}^3 \frac{b_n}{\Lambda_3^{3n}} \left(\partial\pi\right)^2
\!e_{(n)}(\Pi)\nonumber\\ & + \frac{a_3}{\Lambda_3^{6}}
\hat{h}^{\mu\nu} P_{\mu\nu}^{(3)}(\Pi)\,.
\end{align}
We see in the first line the appearance of the ordinary Galileon terms
up to quintic order [we denote
  $(\partial\pi)^2\equiv\partial_\mu\pi\partial^\mu\pi$]. The
coefficients $b_n$ are given by certain combinations of the
$a_n$'s.\footnote{Namely, $b_0=-\frac{3}{4}a_1^2$,
  $b_1=-\frac{3}{4}a_1 a_2$, $b_2=-\frac{1}{2}a_2^2-\frac{1}{3}a_1
  a_3$ and $b_3=-\frac{5}{4}a_2 a_3$.} The last term of
Eq.~\eqref{LgravDLhat} is the remaining mixing between the helicity-0
and helicity-2 modes and is not removable by any local field
redefinition like in~\eqref{fieldredef}.

The contribution of the gravitational sector to the equation of motion
of the helicity-2 field gives
\begin{equation}\label{dLdh}
\frac{\delta\mathcal{L}^\text{DL}_\text{grav}}{\delta
  \hat{h}^{\mu\nu}} = -2 \mathcal{E}^{\rho\sigma}_{\mu\nu}
\hat{h}_{\rho\sigma} + \frac{a_{3}}{\Lambda_3^{6}}
P^{(3)}_{\mu\nu}(\Pi) \,,
\end{equation}
while its contribution to the equation of motion of the helicity-0
field reads
\begin{equation}\label{dLdpi}
\frac{\delta\mathcal{L}^\text{DL}_\text{grav}}{\delta \pi} = -2
\sum_{n=1}^4 \frac{n b_{n-1}}{\Lambda_3^{3(n-1)}} e_{(n)}(\Pi) +
\frac{a_3}{\Lambda_3^{6}} Q_{\mu\nu}^{(2)\rho\sigma}\!(\Pi)
\partial_{\rho}\partial_{\sigma}\hat{h}^{\mu\nu}\,,
\end{equation}
where we posed
\begin{equation}\label{Qeq}
Q_{\mu\nu}^{(2)\rho\sigma}\!(\Pi) \equiv \frac{\partial
  P_{\mu\nu}^{(3)}}{\partial \Pi_{\rho\sigma}} = -\frac{1}{2}
\varepsilon_{\mu}^{\phantom{\mu}\rho\epsilon\lambda}
\varepsilon_{\nu\phantom{\sigma}\pi\tau}^{\phantom{\nu}\sigma}
\Pi^{\pi}_{\epsilon}\Pi^{\tau}_{\lambda}\,.
\end{equation}
The second-order nature of the equations of motion in the gravity
sector is apparent. This is the standard property of the ghost-free
massive gravity interactions~\cite{deRham10, deRham:2010tw}.

\subsection{Matter sector}

As next, we shall control the contributions in the matter sector due
to both the helicity-0 and helicity-2 fields. To this aim it is
important to properly identify the matter degrees of freedom that are
metric independent. These are provided by the coordinate densities
defined as $\rho^*_g = \sqrt{-g}\rho_g u^0_g$ and $\rho^*_f =
\sqrt{-f}\rho_f u^0_f$, and by the ordinary (coordinate) velocities
$v^\mu_g=u^\mu_g/u^0_g$ and $v^\mu_f=u^\mu_f/u^0_f$. The associated
currents $J^{*\mu}_g= \rho^*_g v^\mu_g$ and $J^{*\mu}_f= \rho^*_f
v^\mu_f$ are conserved in the ordinary sense, $\partial_\mu
J^{*\mu}_g=0$ and $\partial_\mu J^{*\mu}_f=0$, and are related to the
classical currents by
\begin{equation}\label{currents*}
J^{*\mu}_g = \sqrt{-g}\,J_g^\mu\quad\text{and}\quad J^{*\mu}_f =
\sqrt{-f}\,J_f^\mu\,.
\end{equation}
When varying the action we must carefully impose that the independent
matter degrees of freedom are the metric independent currents
$J^{*\mu}_g$ and $J^{*\mu}_f$. After variation we may restore the
manifest covariance by going back to the classical currents
using~\eqref{currents*}.

Next we must specify how the matter variables will behave in the DL
when we take the scaling
limits~\eqref{scalingDL1}--\eqref{scalingDL2}. In the DL we want to
keep intact the coupling between the helicity-2 mode $h_{\mu\nu}$ and
the particles living in the $g$ sector, therefore we impose
\begin{equation}\label{scalingT}
T_\text{bar}^{\mu\nu} = M_g
\hat{T}_\text{bar}^{\mu\nu}\quad\text{and}\quad T_g^{\mu\nu} = M_g
\hat{T}_g^{\mu\nu}\,,
\end{equation}
with $\hat{T}_\text{bar}^{\mu\nu}$ and $\hat{T}_g^{\mu\nu}$ remaining
constant in the DL. As for the $f$ particles, \vspace{-0.1cm} in a
similar way we demand that $T_f^{\mu\nu} = M_f \hat{T}_f^{\mu\nu}$
with $\hat{T}_f^{\mu\nu}$ being constant.

The next important point concerns the internal vector field
$\mathcal{A}_\mu$. As we have seen this vector field is a
graviphoton~\cite{SCHERK1979265}, \textit{i.e.} its scale is given by
the Planck mass, witness the factor $M_\text{eff}^2$ in front of the
kinetic term of the vector field~\eqref{W(X)lead}, see also the factor
$M_\text{eff}^2$ in front of the stress-energy tensor of the vector
field, Eq.~\eqref{Tgeff}. For the model to work $M_\text{eff}$ must be
of the order of the Planck mass, as determined
in~\eqref{Mefflambda}. This means that we have to canonically
normalize the vector field $\mathcal{A}_\mu$ according to
\begin{equation}\label{scalingA}
\mathcal{A}_\mu = \frac{\hat{\mathcal{A}}_\mu}{M_\text{eff}}\,,
\end{equation}
and keep $\hat{\mathcal{A}}_\mu$ constant in the DL. Thus
$T_{g_\text{eff}}^{\mu\nu} = \hat{T}_{g_\text{eff}}^{\mu\nu}$ should
be considered constant in that limit.

A general variation of the matter action with respect to the two
metrics reads
\begin{align}
\delta\mathcal{L}_\text{mat} &= \frac{\sqrt{-g}}{2}
\bigl(T_\text{bar}^{\mu\nu} + T_g^{\mu\nu}\bigr)\delta g_{\mu\nu} +
\frac{\sqrt{-f}}{2} T_f^{\mu\nu}\delta f_{\mu\nu} \nonumber\\ &+
\frac{\sqrt{-g_\text{eff}}}{2} T_{g_\text{eff}}^{\mu\nu}\delta
g^\text{eff}_{\mu\nu}\,.
\end{align}
We insert Eqs.~\eqref{eq:fDL}--\eqref{effective} and change the
helicity-2 variable according to~\eqref{fieldredef} to obtain the
contribution of the matter action to the field equation for the
helicity-2 field (in guise $\hat{h}_{\mu\nu}$) as
\begin{align}\label{dLmatdh}
\frac{\delta\mathcal{L}_\text{mat}}{\delta \hat{h}_{\mu\nu}} &=
\frac{1}{M_g}\sqrt{-g} \,\bigl(T_\text{bar}^{\rho(\mu} +
T_g^{\rho(\mu}\bigr)\biggl(\delta^{\nu)}_\rho +
\frac{h^{\nu)}_\rho}{M_g}\biggr)\nonumber\\ & +
\frac{\alpha}{M_g}\sqrt{-g_\text{eff}}
\,T_{g_\text{eff}}^{\rho(\mu}\Bigl(\left(\alpha+\beta\right)\delta^{\nu)}_\rho
+ K^{\nu)}_\rho\Bigr)\,.
\end{align}
Taking the DL with the postulated
scalings~\eqref{scalingT}--\eqref{scalingA} we find that the
helicity-2 mode of the massive graviton is just coupled in this limit
to the baryons and $g$ particles,
\begin{align}\label{dLmatdhDL}
\frac{\delta\mathcal{L}^\text{DL}_\text{mat}}{\delta \hat{h}_{\mu\nu}}
&= \hat{T}_\text{bar}^{\mu\nu} + \hat{T}_g^{\mu\nu} \,,
\end{align}
where the (rescaled) stress-energy tensors
$\hat{T}_\text{bar}^{\mu\nu}$ and $\hat{T}_g^{\mu\nu}$ in the DL are
computed with the Minkowski background.

We next consider the contributions of the matter sector to the
equation of motion of the helicity-0 field. We find three
contributions, two coming from the field
redefinition~\eqref{fieldredef},
\begin{subequations}\label{dLmatdpi1}
\begin{align}
&\frac{\delta\mathcal{L}_\text{mat}}{\delta \pi}{\bigg|}_{(1a)} =
  \frac{a_1}{2M_g}\sqrt{-g} \,\bigl(T_\text{bar}^{\mu\nu} +
  T_g^{\mu\nu}\bigr)\biggl(\eta_{\mu\nu} +
  \frac{h_{\mu\nu}}{M_g}\biggr) \nonumber\\ & \quad\quad +
  \frac{a_2}{M_g \Lambda_3^3}\partial_\nu\biggl[\sqrt{-g}
    \,\bigl(T_\text{bar}^{\mu(\nu} +
    T_g^{\mu(\nu}\bigr)\biggl(\delta_{\mu}^{\rho)} +
    \frac{h_{\mu}^{\rho)}}{M_g}\biggr)\partial_\rho\pi\biggr]
  \,,\label{dLmatdpi1a}\\
%%%%%%%%%%%%%%%%%%%%%%%%%%%%%%%%%%%%%%%%%%%%%%%%%%%%%
&\frac{\delta\mathcal{L}_\text{mat}}{\delta \pi}{\bigg|}_{(1b)} =
  \frac{\alpha a_1}{2M_g}\sqrt{-g_\text{eff}}
  \,T_{g_\text{eff}}^{\mu\nu}\Bigl(\left(\alpha+\beta\right)\eta_{\mu\nu}
  + K_{\mu\nu} \Bigr) \nonumber\\ & \quad\quad + \frac{\alpha a_2}{M_g
    \Lambda_3^3}\partial_\nu\biggl[\sqrt{-g_\text{eff}}
    \,T_{g_\text{eff}}^{\mu(\nu}\Bigl(\left(\alpha+\beta\right)\delta^{\rho)}_\mu
    + K^{\rho)}_\mu\Bigr)\partial_\rho\pi\biggr] \,,\label{dLmatdpi1b}
\end{align}
\end{subequations}
and the third one being ``direct'', and already investigated
in~\cite{Blanchet:2015bia} with result
\begin{align}\label{dLmatdpi2}
\frac{\delta\mathcal{L}_\text{mat}}{\delta \pi}{\bigg|}_{(2)} =&
-\frac{1}{\Lambda_3^3}\partial_\mu\partial_\nu\biggl[\!\sqrt{-f}
  \,T_f^{\rho\mu}\biggl(\delta^\nu_\rho-\frac{\Pi^\nu_\rho}{\Lambda_3^3}
  \biggr) \\ & \quad + \beta \sqrt{-g_\text{eff}}
  \,T_{g_\text{eff}}^{\rho\mu}
  \Bigl(\left(\alpha+\beta\right)\delta^{\nu}_\rho +
  K^{\nu}_\rho\Bigr)\biggr]\,.\nonumber
\end{align}
The latter contribution might look worrisome in the DL, but it becomes
finite after using the equation of motion for the $f$ particles,
Eq.~\eqref{EOMdmf}, and that for the vector field,
Eq.~\eqref{divtaueff}. The calculation proceeds similarly to the one
using Eqs.~(3.29)--(3.32) in Ref.~\cite{dRHRa}. Finally the result can
be brought into the form~\cite{Blanchet:2015bia}
\begin{align}\label{eqpiMatter}
\frac{\delta\mathcal{L}_\text{mat}}{\delta \pi}{\bigg|}_{(2)} &=
\frac{1}{\Lambda_3^3}\partial_\nu\biggl[
  J^{*\rho}_f\,\mathcal{F}_{\mu\rho}\biggl(\eta^{\mu\nu} -
  \frac{\Pi^{\mu\nu}}{\Lambda_3^3}\biggr)^{-1} \\ &+
  \beta\left(J^{*\rho}_g - J^{*\rho}_f\right)
  \mathcal{F}_{\mu\rho}\Bigl(\left(\alpha+\beta\right)\eta^{\mu\nu} +
  K^{\mu\nu}\Bigr)^{-1} \biggr] \nonumber\,,
\end{align}
where we describe the matter degrees of freedom by means of the
coordinate currents~\eqref{currents*}.

The results~\eqref{dLmatdpi2} and~\eqref{eqpiMatter} are general at
this stage, and involve couplings between both the helicity-0 and
helicity-2 modes with the matter fields --- $g$ and $f$ particles, and
the internal vector field $\mathcal{A}_\mu$. However, because of the
scaling~\eqref{scalingA}, which we recall is appropriate to the
graviphoton whose coupling scale is given by the Planck mass, the
vector field strength actually scales like
$\mathcal{F}_{\mu\nu}=\hat{\mathcal{F}}_{\mu\nu}/M_\text{eff}$ in the
DL limit. This fact kills all the interactions between the helicity-0
mode and the vector field in the DL, since they come with an inverse
power of $M_\text{eff}$.\footnote{Note that if we do not impose the
  scaling
  $\mathcal{F}_{\mu\nu}=\hat{\mathcal{F}}_{\mu\nu}/M_\text{eff}$ the
  equation~\eqref{dLmatdh} for the helicity-2 field diverges in the
  DL. Similarly for Eq.~\eqref{dLmatdpi1b}.} Thus the direct
contribution~\eqref{eqpiMatter} is identically zero in the DL, and
only the contribution~\eqref{dLmatdpi1a} is surviving,
while~\eqref{dLmatdpi1b} is also zero. After further simplification
with the matter equations of motion, we obtain (with
$\hat{T}_\text{bar}$ and $\hat{T}_g$ denoting the Minkowskian traces)
\begin{equation}\label{dLmatdpi3}
\frac{\delta\mathcal{L}^\text{DL}_\text{mat}}{\delta \pi} =
\frac{a_1}{2}\,\bigl(\hat{T}_\text{bar} + \hat{T}_g\bigr) +
\frac{a_2}{\Lambda_3^3}\,\bigl(\hat{T}_\text{bar}^{\mu\nu} +
\hat{T}_g^{\mu\nu}\bigr)\partial_\mu\partial_\nu\pi\,.
\end{equation}

Recapitulating, we find that the DL of the model consists of the
following equation for the helicity-2 mode, \textit{i.e.}
$\delta\mathcal{L}^\text{DL}/\delta \hat{h}_{\mu\nu}=0$ or
equivalently
\begin{align}\label{eomh}
-2 \mathcal{E}_{\rho\sigma}^{\mu\nu} \hat{h}_{\rho\sigma} +
\frac{a_{3}}{\Lambda_3^{6}} P_{(3)}^{\mu\nu}(\Pi) +
\hat{T}_\text{bar}^{\mu\nu} + \hat{T}_g^{\mu\nu} = 0\,,
\end{align}
which is of second-order nature. Thus, the contributions of the
gravitational and matter sector to the equations of motion of the
helicity-2 mode in the DL are ghost-free. Note, that the Bianchi
identity of this equation (taking the divergence of it) is identically
satisfied, since the particles actually follow geodesics in the
DL. Indeed, using~\eqref{scalingT}--\eqref{scalingA} together with the
equations of motion [\textit{e.g.}~\eqref{EOMdm}], we have
$\partial_\nu\hat{T}_\text{bar}^{\mu\nu}=\partial_\nu\hat{T}_g^{\mu\nu}=0$
(the particles move on Minkowski straight lines).

In addition we have the total equation of motion of the helicity-0
mode, namely $\delta\mathcal{L}^\text{DL}/\delta \pi=0$ which reads
\begin{align}\label{eompi}
&\hspace{-0.5cm}-2 \sum_{n=1}^4 \frac{n b_{n-1}}{\Lambda_3^{3(n-1)}}
  e_{(n)}(\Pi) + \frac{a_3}{\Lambda_3^{6}}
  Q_{\mu\nu}^{(2)\rho\sigma}\!(\Pi)
  \partial_{\rho}\partial_{\sigma}\hat{h}^{\mu\nu} \\& \qquad = -
  \frac{a_1}{2}\,\bigl(\hat{T}_\text{bar} + \hat{T}_g\bigr) -
  \frac{a_2}{\Lambda_3^3}\,\bigl(\hat{T}_\text{bar}^{\mu\nu} +
  \hat{T}_g^{\mu\nu}\bigr)\partial_\mu\partial_\nu\pi \,.\nonumber
\end{align}
Since this equation is perfectly of second-order in the derivatives of
the $\pi$ field, we conclude our study by stating that the model is
safe (ghost-free) up to the strong coupling scale. Below that scale
the theory is perfectly acceptable as an Effective Field Theory, and
its consequences can be worked out using perturbation theory as
usual. For instance, solving at linear order the helicity-0
equation~\eqref{eompi} we obtain the usual well-posed
(hyperbolic-like) equation
\begin{equation}\label{hyperbolic}
\Box\pi = \frac{a_1}{4b_0}(\hat{T}_\text{bar} +
\hat{T}_g)+\mathcal{O}\left(\pi^2\right)\,,
\end{equation}
which can then be perturbatively iterated to higher order. With this
we have proved, that the coupling of the dark matter particles with
the internal vector field does not introduce any ghostly contribution
in the DL.

%%%%%%%%%%%%%%%%%%%%%%
%%%%%%%%%%%%%%%%%%%%%
\section{Conclusions}
\label{sec:concl}

This work was dedicated to the detailed study of the decoupling limit
interactions of the dark matter model proposed
in~\cite{Blanchet:2015sra, Blanchet:2015bia}. This model is
constructed \textit{via} a specific coupling of two copies of dark
matter particles to two metrics in the framework of massive
bigravity. Furthermore, an internal vector field links the two dark
matter species. This enables us to implement a mechanism of
gravitational polarization, which induces the MOND phenomenology on
galactic scales (with the specific choice of parameters studied
in~\cite{Blanchet:2015bia}). Note that, since our model successfully
reproduces all aspects of that phenomenology, it will be in agreement
with the recent observations of the MOND mass-discrepancy-acceleration
relation in~\cite{PhysRevLett.117.201101}.

Some theoretical and phenomenological consequences of this model were
studied in detail in Ref.~\cite{Blanchet:2015bia}, but it was also
pointed out that the decoupling limit of the theory may be
problematic, with higher derivative terms occuring in the equation of
motion of the helicity-0 mode of the massive graviton.

In the present work, we studied the complete DL interactions crucially
including the contributions of the matter sector, and we showed that
by necessary rescaling of the vector field (as appropriate for a
vector field with Planckian coupling constant) the theory is free from
ghosts in the DL, and hence can be used as a valid Effective Field
Theory up to the strong coupling scale.

\vspace{0.3cm}
\acknowledgments
\vspace{-0.3cm} 

We would like to thank Claudia de Rham and Andrew Tolley for very
useful and enlightening discussions. L.H. wishes to acknowledge the
Institut d'Astrophysique de Paris for hospitality and support at the
final stage of this work.

\bibliography{DDM_LB_LH.bib}

\end{document}